\documentclass[11pt]{article}
\usepackage{moriond,epsfig}

\def\Journal#1#2#3#4{{#1} {\bf #2}, #3 (#4)}

\def\NPA{{\em Nucl. Phys.} A}

\def\PLB{{\em Phys. Lett.}  B}
\def\PRL{\em Phys. Rev. Lett.}

\def\PRC{{\em Phys. Rev.}  C}
\def\PRD{{\em Phys. Rev.} D}

\def\EPJC{{\em Eur. Phys. J.} C}
\def\JPG{{\em J. Phys.} G}

\def\be{\begin{equation}}
\def\ee{\end{equation}}
\def\bea{\begin{eqnarray}}
\def\eea{\end{eqnarray}}

\begin{document}
\vspace*{4cm}
\title{J/$\psi$ production measurements by the PHENIX experiment at RHIC}

\author{ Ermias T. Atomssa, for the PHENIX collaboration }

\address{Laboratoire Leprince Ringuet, \'Ecole Polytechnique/IN2P3,
  \\91128, Palaiseau, France } 

\maketitle\abstracts{
Measurements of J/$\psi$ production by the PHENIX experiment in p+p,
d+Au, Cu+Cu and Au+Au collisions at $\sqrt{s_{NN}}$=200~GeV are
reviewed. The results show a suppression beyond what can be explained
by cold nuclear matter effects in the most central Au+Au and to a
lesser extent in Cu+Cu collisions. In addition, the suppression
observed at mid rapidity in Au+Au is smaller than at forward rapidity,
a tendency opposite to what is expected from the higher energy density
at mid rapidity. Regeneration, a possible explanation, can be tested
by measuring the elliptic flow parameter v$_2$ of J/$\psi$.}

\section{Introduction}
J/$\psi$ suppression is considered to be one of the key probes of the Quark
Gluon Plasma (QGP) formation in heavy ion collisions. Color screening
was proposed~\cite{satz} as a mechanism leading to {\it anomalous }
suppression beyond {\it normal} hadronic absorption if J/$\psi$s are
created in a deconfined medium. The CERN SPS experiments NA38, NA50
and NA60 were the first to investigate this phenomenon by measuring
J/$\psi$ suppression in a variety of colliding systems and energies. The
results show a statistically significant anomalous
suppression in central Pb+Pb~\cite{spsqgp} and In+In~\cite{spsqgp2}
collisions, that can be interpreted in terms of melting in the QGP.

\begin{figure}
\begin{center}
  \begin{tabular}{cc}
  \includegraphics[height=4.5cm]{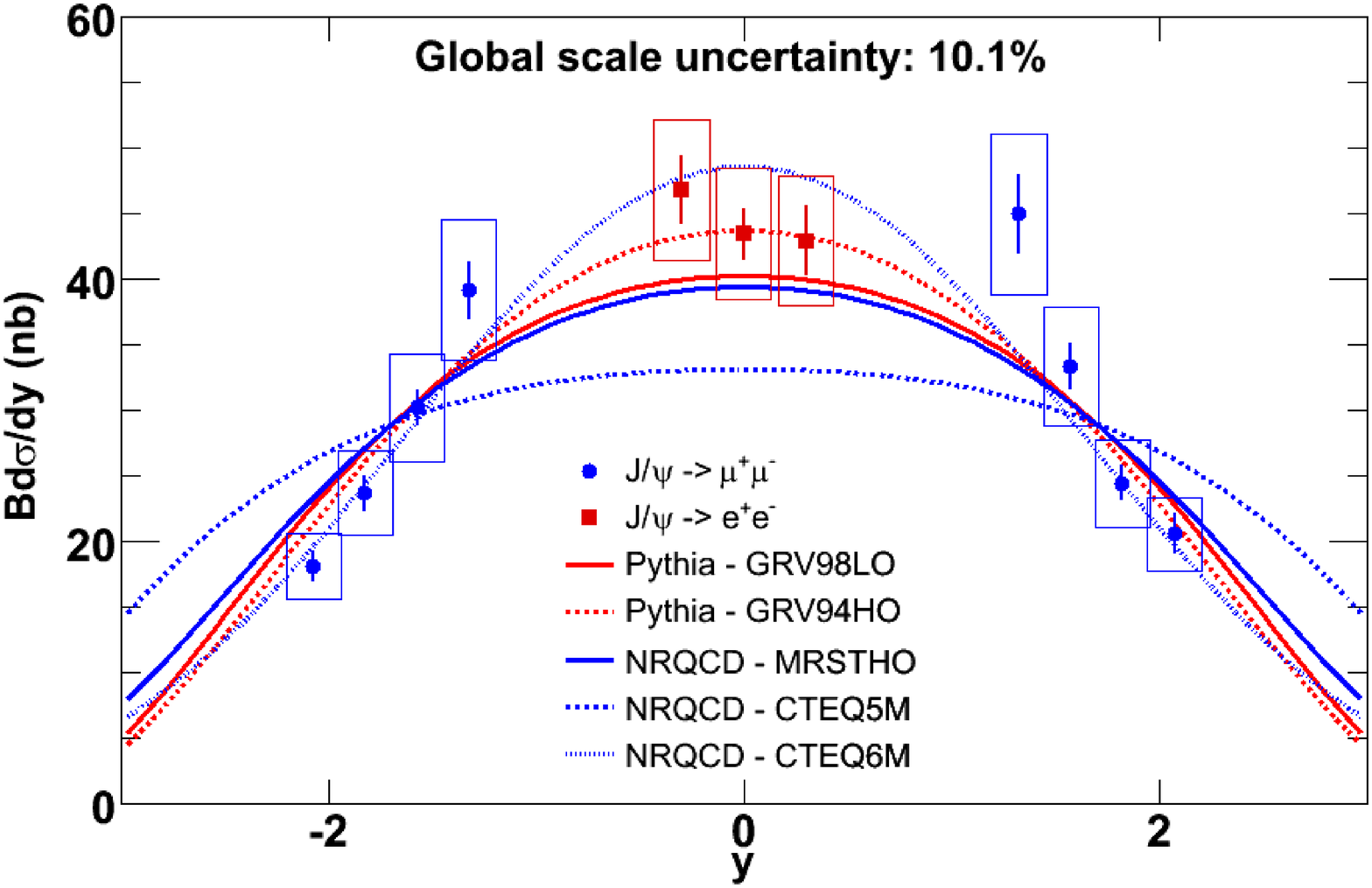}
  & \includegraphics[height=4.5cm]{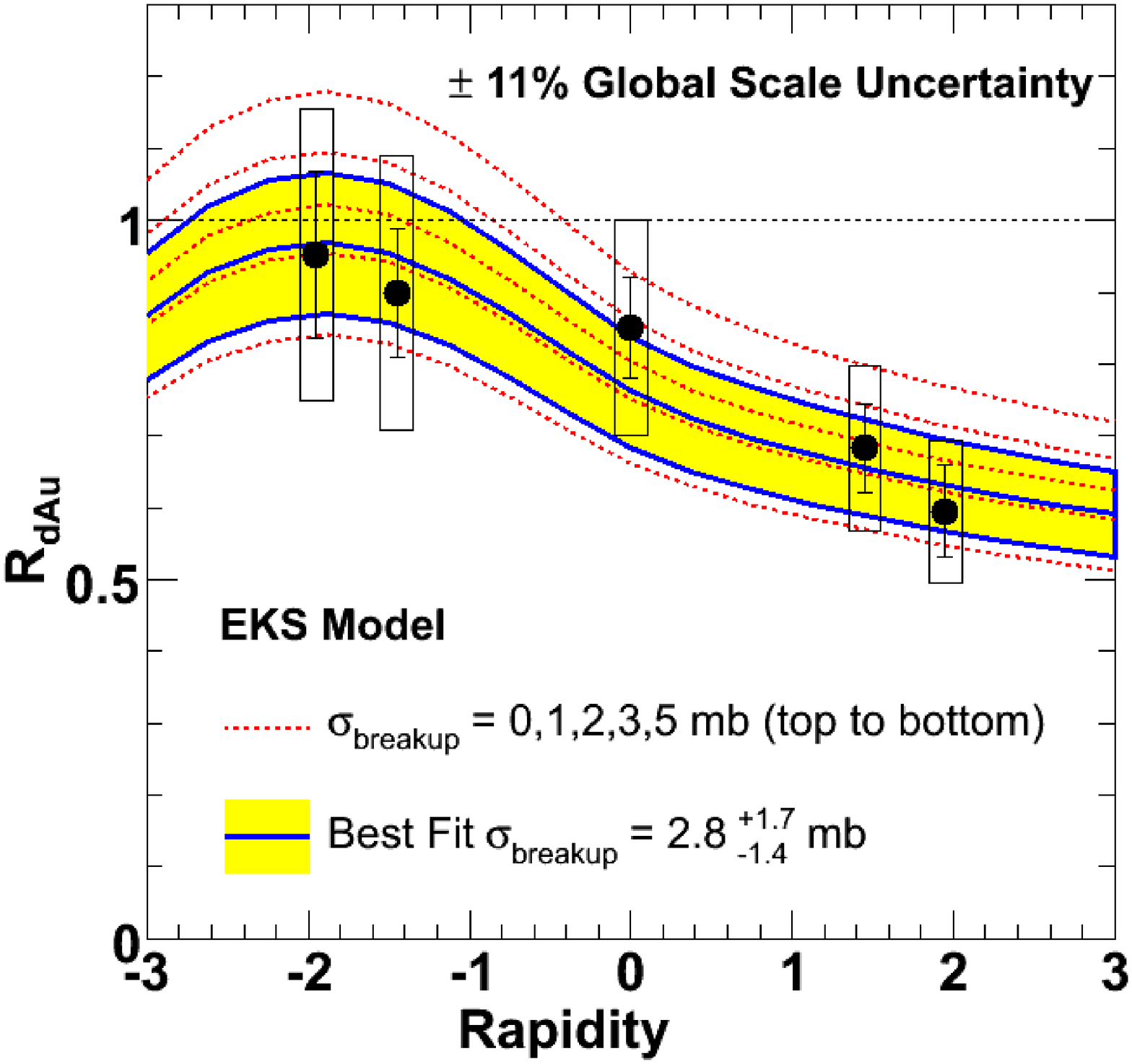}
  \end{tabular}
\end{center}
\vspace{-2ex} 
\caption{ J/$\psi$ cross section vs. rapidity in p+p collisions
  (left). J/$\psi$ R$_{dA}$ in d+Au collisions vs. rapidity (right).
\label{fig:1}}
\vspace{-2ex}
\end{figure}

The PHENIX experiment at RHIC has also measured the production of
J/$\psi$ in a variety of colliding systems, and provided further
insights by exploring this phenomenon at higher energies. J/$\psi$s
are detected in PHENIX through their dielectron decay at mid rapidity
($|y|<0.35$) and through their dimuon decay at forward rapidity
($1.2<|y|<2.4$). J/$\psi$ suppression is characterized by a ratio
called the {\it nuclear modification factor}, obtained by normalizing the
J/$\psi$ yields in heavy ion collisions ($dN_{AB}$) by the J/$\psi$
yields in p+p collisions at the same energy ($dN_{pp}$) times the
average number of binary inelastic nucleon-nucleon collisions
($<N_{coll}>$):
\be
R_{AB} (y, p_{T}) = \frac{ dN_{AB}(y, p_{T})/dydp_{T}} { <N_{coll}>
  dN_{pp}(y,p_{T})/dydp_{T}}.
\label{eq:raa}
\ee
If the heavy ion collision is a superposition of independent N$_{coll}$
inelastic nucleon-nucleon collisions, R$_{AB}$ will be equal to unity,
whereas it will be larger than one in case of enhancement and lower
than one in case of suppression.

At typical RHIC energies ($\sqrt{s_{NN}}=$~19.6~-~200~GeV), J/$\psi$s are
dominantly produced through gluon fusion. The J/$\psi$ yield is
therefore sensitive to gluon shadowing~\footnote{
Shadowing refers to
  the depletion of low momentum partons in nucleons bound
  in nuclei as compared to free nucleons.}. Part of the
ground state charmonia yield also comes from feed down of excited
states ($\psi'$ and $\chi_c$), and can contribute up to
$\sim$~40\% to the total J/$\psi$ yield. In subsequent stages of the
collision involving heavy ions, there are a number of competing
mechanisms that can enhance or suppress the J/$\psi$ yield. The two
major contributors to the suppression are absorption by nuclear
fragments from incident nuclei, and an eventual melting in the
QGP. Finally it is not impossible that a pair of uncorrelated $c$ 
and $\bar{c}$ quarks that are close enough in phase space recombine to
form a bound charmonium state and enhance the J/$\psi$ yield.

\section{Baseline and cold nuclear matter effect measurements }

The differential cross section of J/$\psi$ in p+p collisions as a
function of rapidity measured by PHENIX is shown in
Fig.~\ref{fig:1}~(left)~\cite{pp}. In addition to providing normalization
cross sections essential for the calculation of R$_{AA}$ as in
Eq. \ref{eq:raa}, J/$\psi$ measurements in p+p collisions constrain the
poorly understood J/$\psi$ production mechanism. In the same figure,
predictions from various LO and NLO calculations are shown. The current
precision does not discriminate between the models, but there is
potential for improvement through better precision in cross section
and additional information from J/$\psi$ polarization measurements.

Nuclear absorption and shadowing, collectively referred to as {\it
  cold nuclear matter effects (CNM) } can be constrained by
measurements in proton (or light ion) on heavy ion collisions. In
PHENIX this was performed in deuteron-gold collisions. The resulting 
suppression ratio R$_{dAu}$ is shown in Fig.~\ref{fig:1}~(right)~\cite{dau} as a
function of rapidity, where the positive rapidity coincides with the
deuteron going direction. J/$\psi$s detected in different rapidity
regions probe specific gluon x$_2$ regions~\footnote{
By x$_2$,
  we refer to the parton longitudinal momentum fraction in the
  nucleus.}. Forward rapidity corresponds to x$_{2}$
$\sim$~0.002~-~0.01 where the depletion due to shadowing is important
whereas backward rapidity corresponds to x$_{2}$
$\sim$~0.05~-~0.2 where a slight enhancement due to
anti-shadowing is expected. The rapidity dependence of R$_{dA}$ therefore
reflects the gluon shadowing, whereas the global vertical scale
is determined by the amount of {\it normal} absorption.

To quantitatively disentangle the shadowing component from the
absorption component, a rapidity dependence using two shadowing
schemes, EKS~\cite{eks} and NDSG~\cite{ndsg} was fitted to R$_{dA}$
leaving the overall vertical scale a free parameter to account for the
absorption~\cite{dau}. J/$\psi$ absorption cross sections of
$2.8 ^{+1.7}_ {-1.4}$~mb and $2.2^{+1.6}_{-1.5}$~mb
were obtained for EKS and NDSG schemes respectively. This is in
agreement with the absorption cross section 
reported by the SPS of $4.2 \pm 0.5$~mb~\cite{sps}
but such a comparison should not be taken at face value because
shadowing is not taken into account in the SPS~\footnote{
Taking
  into account nuclear PDF modification would increase the SPS
  absorption cross section, because the SPS rapidity corresponds to
  the anti-shadowing regime, requiring more absorption to
  account for the observed suppression.} absorption cross sections
evaluation.

\section{Anomalous suppression in heavy ion systems}
\begin{figure}[htbp]
\begin{center}
  \begin{tabular}{cc}
  \includegraphics[height=4.5cm]{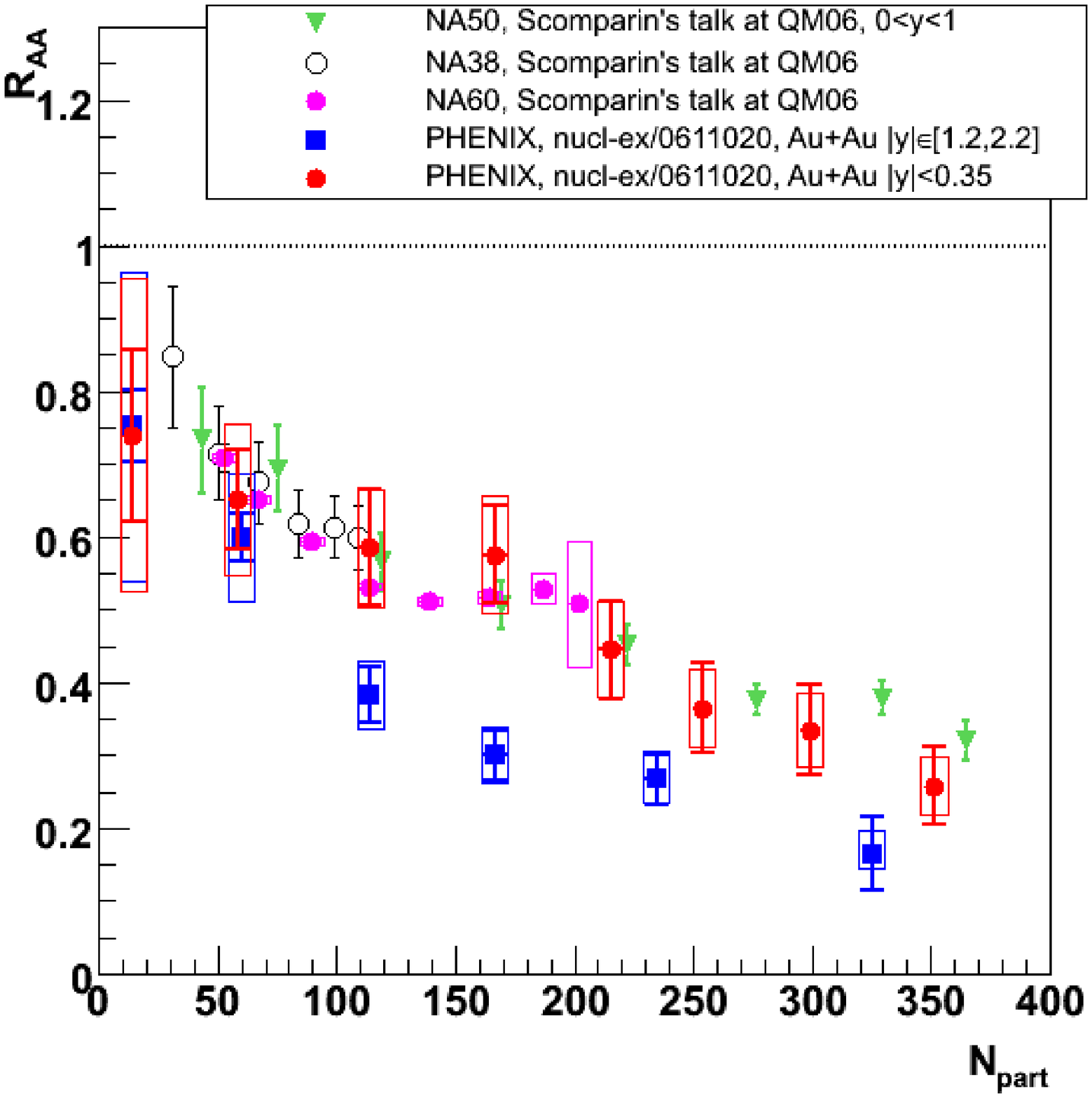}
  & \includegraphics[height=4.5cm]{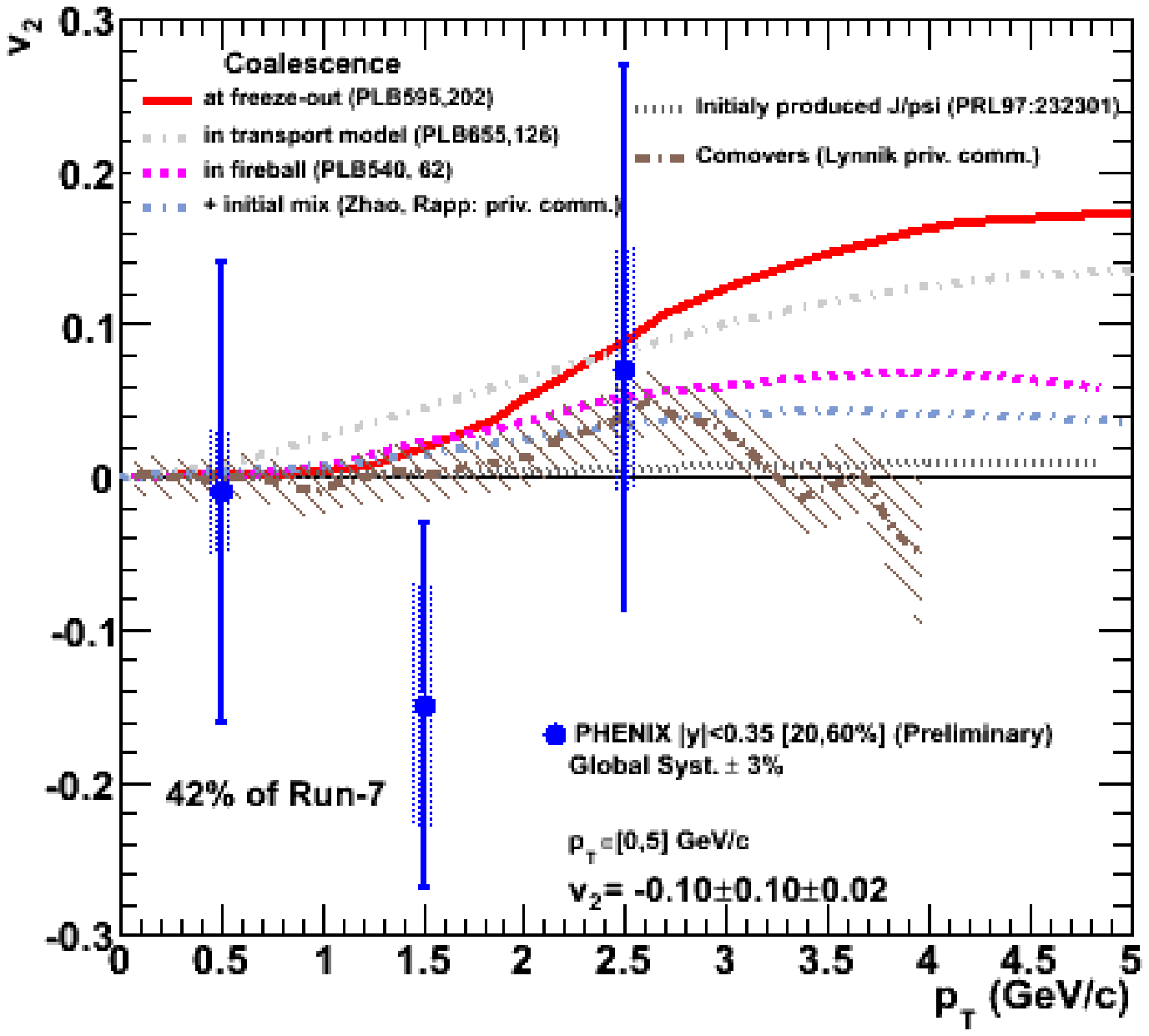}
  \end{tabular}
\end{center}
\vspace{-2ex} 
\caption{ J/$\psi$ R$_{AA}$ vs. N$_{part}$ at SPS compared to
  RHIC (left). J/$\psi$ v$_2$ vs. p$_T$ measurement by PHENIX (right).\label{fig:3}}
\vspace{-2ex}
\end{figure}

PHENIX has also measured J/$\psi$ suppression in Au+Au~\cite{auau} and
Cu+Cu~\cite{cucu} collisions at $\sqrt{s_{NN}} = 200$~GeV. The
J/$\psi$ R$_{AA}$ in Au+Au collisions as a function of the number of
participants N$_{part}$, at forward and mid rapidity ranges is shown in
Fig.~\ref{fig:3}~(left)~\cite{auau} together with data points from NA38,
NA50 and NA60 experiments. The R$_{AA}$ goes down to $\sim$~0.2
for the most central Au+Au collisions (large N$_{part}$), and
approaches unity for peripheral ones (small N$_{part}$). To see the
extent of anomalous suppression, extrapolations of the CNM and
shadowing constraints obtained from d+Au measurements were calculated
using a model dependent method which assumes the above mentioned
shadowing schemes as well as with a data driven method which has
minimal model dependence. The result~\cite{dau} from both methods is a
statistically significant suppression beyond CNM extrapolation in the
most central forward rapidity Au+Au collisions, less pronounced at mid
rapidity Au+Au or in Cu+Cu collisions.

The data show two features that contradict local density induced
suppression models. The mid
rapidity suppression is lower than the forward rapidity suppression
(cf.~Fig.~\ref{fig:3}~(left)), despite experimental evidence~\footnote{
The rapidity density of charged particles which 
  increases with the deposited energy peaks at mid
  rapidity~\cite{phobos}. } that energy density is higher at mid
rapidity than at forward rapidity. The same remark holds for the
comparison between R$_{AA}$ at mid rapidity in PHENIX and R$_{AA}$
at SPS~\footnote{
Care must be taken when comparing with SPS, because the CNM effects
are not the same at the two energies.} 
(cf.~Fig.~\ref{fig:3}~(left)). The two are in agreement within error bars, a
surprising result considering that the energy density reached at RHIC
is larger than the one reached at SPS. A number of explanations have
been put forth, including sequential melting, where only $\psi$' and
$\chi_c$ are dissociated leading to a suppression of only the
feed down component of the $J/\psi$ yield~\cite{kks}, and gluon
saturation that leads to a lower charm quark yield at forward
rapidity~\cite{tuchin}.

\section{Regeneration}

A strong regeneration of J/$\psi$ from uncorrelated $c$ and $\bar{c}$
quarks is another good candidate to explain the tendency of R$_{AA}$ as a
function of rapidity at RHIC. This is supported by the high charm quark yield
measurements~\cite{openc} ($\sim$10 $c\bar{c}$ pairs are created in
the most central Au+Au collisions). A number of model predictions that
incorporate regeneration have been proposed~\cite{regen} and all of
them reproduce qualitatively the rapidity dependence of J/$\psi$
R$_{AA}$ observed by PHENIX.

However, important inputs to regeneration models such as the precise number of
$c$$\bar{c}$ pairs available for recombination and the phase space
conditions for recombination to take place are poorly constrained. It
is thus very compelling to have a direct experimental check 
of regeneration. The J/$\psi$ elliptic flow is one candidate. Elliptic
flow refers to the azimuthal angle correlation of  particle emission
with respect to the reaction plane orientation~\footnote{
The reaction
  plane is the plane defined by the beam axis and the line joining the
  center of colliding nuclei. It is measured in PHENIX from azimuthal
  angle distribution of charged particles close to beam rapidity.}. It
is quantified by the second Fourier coefficient v$_2$ of the azimuthal
angle distribution of identified particles. The measured v$_2$ of
electrons from D and B meson decays is remarkably
high~\cite{flowc}. This is believed to originate from the elliptic
flow of underlying charm and beauty quarks. J/$\psi$s from
recombination should inherit the charm quark flow, resulting in a
higher v$_2$ than the case of direct production in hard collisions.

The first measurement of J/$\psi$ v$_2$ at RHIC energy was performed by
PHENIX at mid rapidity and is shown in Fig.~\ref{fig:3}~(right) as a function
of transverse momentum. Predictions from models that assume various
amounts of recombination from none to full coalescence at freeze out
are plotted together. Data points are compatible within the error bars
simultaneously with zero flow as well as with the model that predicts
maximum flow. This result should therefore be seen as {\it proof of
  principle} of the feasibility of J/$\psi$ v$_2$ measurements. There
is still room for improvement using already existing data,
but it is to be noted that a much larger sample will probably be
needed to be able to distinguish between the different models.


\section*{References}
\begin{thebibliography}{99}
\bibitem{satz} T. Matsui and H. Satz, \Journal{\PLB}{178} {416}{1986} 
\bibitem{spsqgp} B. Alessandro {\it et al.},
  \Journal{\EPJC}{39}{335-345}{2005}
\bibitem{spsqgp2} R. Arnaldi {\it et al.},
  \Journal{\PRL}{99}{132302}{2007} 
\bibitem{pp} A. Adare {\it et al.}, \Journal{\PRL}{98}{232002}{2007} 
\bibitem{dau} A. Adare {\it et al.},
  \Journal{\PRC}{77}{024912}{2008}
\bibitem{eks} K. S. Eskola {\it et al.},
  \Journal{\NPA}{696}{729}{2001} 
\bibitem{ndsg} D. deFlorian {\it et al.},
  \Journal{\PRD}{69}{074028}{2004} 
\bibitem{sps} B. Alessandro {\it et al.},
  \Journal{\EPJC}{48}{329}{2006} 
\bibitem{auau} A. Adare {\it et al.},
  \Journal{\PRL}{98}{232301}{2007}
\bibitem{cucu} A. Adare {\it et al.}, nucl-ex/0801.0220 
\bibitem{phobos} B. B. Back {\it et al.},
  \Journal{\PRL}{91}{052303}{2003}
\bibitem{kks} F. Karsch {\it et al.}, \Journal{\PLB}{637}{75}{2006}
\bibitem{tuchin} K. Tuchin, \Journal{\JPG} {30}{S1167-S1170}{2004}
\bibitem{openc} S. S. Adler {\it et al.}, \Journal{\PRL}{94}{082301}{2005}
\bibitem{regen} R. Thews {\it et al.}, \Journal{\EPJC}{43}{97}{2005};
  L. Yan {\it et al.}, \Journal {\PRL}{97}{232301}{2006}; A. Andronic
  {\it et al.}; \Journal{\NPA}{789}{34}{2007}; L. Ravagli and R. Rapp, 
  \Journal{\PLB}{655}{p126}{2007}; X. Zhao and R. Rapp,
  arXiv:0712.2407; K. Tywoniuk {\it et al}, arXiv:0804.4320; O. Linnyk
  {\it et al}, arXiv:0801.4282
\bibitem{flowc} A. Adare {\it et al.} \Journal{\PRL}{98}{172301}{2007}

\end{thebibliography}
\end{document}